\documentclass[12pt]{iopart}

\usepackage{color}

\usepackage{bm}
\usepackage{amssymb}
\usepackage{graphicx}
\usepackage{array}
\usepackage{tabulary}
\usepackage{caption}

\usepackage{etoolbox}
\apptocmd{\sloppy}{\hbadness 10000\relax}{}{}

\newcolumntype{K}[1]{>{\centering\arraybackslash}p{#1}}

\begin{document}

\title{Probing Axial-Vector Charmonia $\chi_{c1}(1P)$ and $\chi_{c1}(2P)$}

\author{J.Y. S\"ung\"u and A.C. Jumasahatov}

\address{Department of Physics, Kocaeli University, 41380
Izmit, Turkey}

\begin{abstract}
We investigate the ground-state heavy quarkonium $\chi_{c1}(1P)$
and its first excited state $\chi_{c1}(2P)$ with quantum numbers
$J^{PC}=1^{++}$. The masses and decay constants of these
charmonium states are computed using two-point QCD sum rule method
by including quark, gluon and mixed condensates up to dimension-8.
We compare our numerical results with the available experimental
data as well as existing theoretical predictions in the
literature.
\end{abstract}
\section{Introduction}

As a bound state of charm-anticharm pair, heavy charmonia is an
ideal testing ground to understand the hadron dynamics and play
conspicuous role in the strong interactions between quarks in the
interplay of perturbative and non-perturbative regime. However,
there are many unresolved questions about charmonia in this
regime. Most strikingly, there are a number of new charmonium
states, called as XYZ particles, that could not be interpreted
thoroughly until now and could not be placed into a
well-established meson groups~\cite{Godfrey:2008nc,Chen:2016qju}.

There has been a great progress in the observation of the
charmonia from the past few years \cite{Tanabashi}. But the higher
$c\overline{c}$ states with $J^{PC}=1^{++}$, such as
$\chi_{c1}(2P), \chi_{c1}(3P)$ and $\chi_{c1}(4P)$ are still not
established. Currently attention is focused on the $\chi_{c1}(2P)$
particle as one of the special charmonia. We have no exact
knowledge on ($2^1P_1$) $\chi_{c1}(2P)$ yet. There are quite
different opinions on $\chi_{c1}(2P)$ state in the literature.
Some of them proposed that it is possible to describe the
$X(3872)$ meson as the first excited state of $\chi_{c1}(1P)$
meson with a little mass shift. But others claim that $X(3872)$
resonance can not be $\chi_{c1}(2P)$ meson \cite{Hanhart:2017kbu}
which conflicts with the prediction of the quark model. Also, in
Ref. \cite{Gui:2018rvv} $X(4274)$ is assigned as the
$\chi_{c1}(3P)$ state. In this case the first question that comes
to our mind is what is the nature of the $X(4140)$. Recently,
$X(4140)$ state is renamed as $\chi_{c1}(4140)$ with $I^G
J^{PC}=0^+(1^{++})$ in PDG \cite{Tanabashi}. Another possibility
for the first excited state of the $\chi_{c1}$ meson is X(3940).
But we do not know the quantum numbers of this state yet
\cite{Abe:2007sya,Abe:2007jna}. These puzzle has been discussed
exhaustively in the literature \cite{Chen:2016qju}, but a
consistent description is still missing \cite{Zhou:2017dwj}.

The bare mass of $\chi_{c1}(2P)$ is found to be $3950$~MeV in the
GI~Model, while its coupling to the $D\bar{D}^*$ threshold reduces
its pole mass to $3884$~MeV depending on the parameter set
selected~\cite{Godfrey:1985xj}. Besides mass for the resonance
$\chi_{c1}(2P)$ is obtained as $3872$ MeV in
Ref.~\cite{DSouza:2017hno}. According to the prediction of the
naive potential model, the mass of $X(3872)$ is $50$~MeV lower
than the mass of $\chi_{c1}(2P)$, too. Also, Achasov and et al.
elucidate the mass shift of $X(3872)$ regarding the estimation of
the potential model for the mass of $\chi_{c1}(2P)$ with the
contribution of the virtual $D\bar{D}^*+c.c.$ intermediate states
into the self energy of $X(3872)$~\cite{Achasov:2015oia}.
Additionally in Ref.~\cite{Wang:2010ej} radiative E1 decay widths
of $X(3872)$ are calculated by the Relativistic Salpeter method,
with the assumption that $X(3872)$ is the pure $\chi_{c1}(2P)$
state and fitted the model parameters for $\chi_{c1}(2P)$.
Presuming $X(3872)$ as the radial excited state of
$\chi_{c1}(1P)$, the ratio of $\mathcal{B}({X(3872) \rightarrow
\psi(2S)\gamma})/\mathcal{B}({X(3872) \rightarrow J/\psi\gamma})$
is found as $4.4$, which is consistent with the experimental
result by BaBar~\cite{Aubert:2008ae}, but is larger than the upper
bound reported by Belle~\cite{Bhardwaj:2011dj}.

Moreover, study of the radiative decays of $X(3872)$ by using
effective Lagrangian approach shows that identification of this
state with $\chi_{c1}(2P)$ is plausible~\cite{DeFazio:2012sg}. In
Ref.~\cite{Li:2009ad}, Li and et al. showed that the S-wave
coupling effect on lowering the $\chi_{c1}(2P)$ mass towards the
$D\bar{D}^*$ threshold supports assignment of the $X(3872)$ as a
pure $\chi_{c1}(2P)$ charmonia. Likewise the authors
of~\cite{Zhou:2013ada} analyze the pole trajectory of the
$\chi_{c1}(2P)$ state while quark pair production rate from the
vacuum changes in its uncertainty region, which denotes that the
enigmatic $X(3872)$ resonance may be defined as a $1^{++}$,
$\bar{c}c$ charmonium dominated state dressed by the hadron loops.
As Anisovich and et al. stated in \cite{Anisovich:2005vs} the
$X(3872)$ can be either $\chi_{c1}(2P)$ state or $\eta_{c2}(1D)$
based on the study of radiative transitions. Using
Friedrichs-Model-like scheme, Zhou and et al. concluded that the
$X(3872)$ could be dynamically generated by the coupling of the
bare resonance $\chi_{c1}(2P)$ and continuums, however its
continuum part is larger. This proposal is encouraging in matching
the prediction of GI Model with observed
states~\cite{Zhou:2017dwj}. As a result, to clarify the situation
on $\chi_{c1}(2P)$ and $X(3872)$ we need to determine hadronic
measurables precisely in experiments and confirm the numerical
values of these parameters with theoretical predictions.

Further numerous theoretical studies on the resonances of
$X(4140)$ and $X(4274)$ have been performed in the literature
treating them as the compact or diquark-antidiquark states,
molecular states, hybrid charmonium states, dynamically generated
resonances, conventional charmonium, and cusp effects
\cite{vanBeveren:2009dc,Swanson:2014tra,Agaev:2017foq,Turkan:2017pil,Stancu:2009ka}.
When these states are assigned as $J^{PC}=1^{++}$, molecular and
hybrid charmonium interpretations with other quantum numbers can
be dismissed. There are possibilities for a non-resonance
interpretation for $Y(4140)$, such as the $D_sD_s^*$ cusp
\cite{Aaij:2016iza,Aaij:2016nsc} or $D_sD_s^*$ re-scattering via
the open-charmed meson loops \cite{Liu:2016onn}. Note that the
cusp effects may explain the structure of the $X(4140)$, but fail
to account for the $X(4274)$ \cite{Swanson:2014tra}. What is more,
the compact tetraquark scenario can describe the $X(4140)$ and
$X(4274)$ simultaneously \cite{Stancu:2009ka}, while only one
$J^{PC} = 1^{++}$ state exists in the color triplet
diquark-antidiquark picture in this energy region
\cite{Ebert:2010zz}.

When fitted as a resonance, its mass
$(4146.5\pm4.5^{+4.6}_{-2.8})$ MeV is in excellent agreement with
earlier measurements for the $X(4140)$, whereas the width
$(83\pm21^{+21}_{-14})$ MeV is substantially larger. The upper
limit previously set for production of a narrow ($15.3$ MeV)
$X(4140)$ based on a small subset of our present data
\cite{Aaij:2012pz}. The $X(4140)$ width is substantially larger
than previously determined \cite{Aaij:2016iza}. In the screened
potential model, the mass of $\chi_{c1}(3P)$ is predicted to be
$4.19$ MeV. Thus, the charmonium-like state X(4140) can be a
candidate for the $\chi_{c1}(3P)$ state. In contrast with the
linear potential model (LP) model calculations, the mass of
$\chi_{c1}(3P)$ is estimated to be $4.28$ MeV which is close to
X(4274). In the LP model, charmonium-like states $X(4274)$ seems
to be a candidate of $\chi_{c1}(3P)$ \cite{Gui:2018rvv}.
Nevertheless larger data samples will be needed to resolve this
issue.

One reliable way for computing hadronic parameters is the QCDSR
technique, which is an analytic formalism steadily established on
QCD and has been successfully applied to many hadrons
~\cite{Shifman1,Shifman2}. In this work, we assume that hadronic
parameters of the first excited state of charmonium state
$\chi_{c1}(1P)$ could be reproduced in a standard QCD sum rule
(QCDSR) calculations subtracting the ground state contribution
from the first excited state and the mass and decay constant of
$\chi_{c1}(2P)$ can be estimated.  During our calculations the
two-point QCDSR is utilized taking into account vacuum condensates
up to dimension-8. Using the relevant currents, the QCDSR have
been obtained and the masses and decay constants of charmonia
$\chi_{c1}(1P)$ and $\chi_{c1}(2P)$ are extracted. So, these
results may be helpful in identifying and completion of the hadron
spectrum at P-wave sector.

The rest of the paper is organized as follows. In
Section~\ref{sec:SR}, we briefly review the basic concepts of the
QCDSR approach used in our calculations. The masses and decay
constants of the heavy axial-vector charmonia $\chi_{c1}(1P)$ and
$\chi_{c1}(2P)$ are derived from QCDSR. Then numerical values are
presented in Section~\ref{sec:NumAnal}. Finally, in
Section~\ref{sec:Result} we compare our results with the findings
of the other models in the literature.

\section{Theoretical Framework\label{sec:SR}}

According to the idea of the QCDSR technique
~\cite{Shifman1,Shifman2}, the short distance perturbative QCD is
extended by the operator product expansion (OPE) of the
correlator, which leads a series in powers of the squared momentum
with Wilson coefficients. The convergence at low momentum or long
distance is improved by imposing Borel transformation. The
quark-based (called as OPE or QCD side) evaluation of the
correlator is equalized to the correlator, computed using hadronic
degrees of freedom (i.e. phenomenological or physical side) via
dispersion relation. Later we obtain the QCDSR from which any
hadronic quantity can be found.

Instead of the approvement that $X(3872)$ state is the $c\bar{c}$
charmonia $\chi_{c1}(2P)$, we considered the scenario where the
$\chi_{c1}(2P)$ is the first excited state of $\chi_{c1}(1P)$. To
extract the hadronic parameters of P-wave ground state and its
first excited state of $\chi_{c1}$, we employ the QCDSR formalism.

In this context, first we determine the sum rules for the mass
$m_{\chi_{c1}}$ and decay constant $f_{\chi_{c1}}$ of the ground
state $\chi_{c1}(1P)$. Then, we use the ``ground state +
continuum'' approximation. Next the ``ground state + first excited
state + continuum'' assumption is used to find the sum rules. So
the masses and decay constants of these mesons can be derived from
these expressions. Obtained numerical values for the ground state
$\chi_{c1}(1P)$ are utilized as input parameters in the sum rules
belonging to the excited one.

According to the QCDSR, hadrons are symbolized by their
interpolating currents and placing the current expression into the
two-point correlator just as creation and annihilation operator
the following expression can be written:
\begin{equation}\label{eq:CorrFunc}
\Pi_{\mu \nu }(q)=i\int d^{4}x~e^{iq\cdot x}\langle
0|\mathcal{T}\big[J_{\mu}^{\chi_{c1}}(x)J_{\nu}^{\dagger
\chi_{c1}}(0)\big]|0\rangle.
\end{equation}
For the meson current $J_{\mu}(x)$ with the $J^{PC}=1^{++}$
following definition is used~\cite{Reinders}:
\begin{equation}\label{eq:Current}
J_{\mu}^{\chi_{c}}(x)=\bar{c}_{i}(x)\gamma_{\mu}\gamma_{5}{c}_{i}(x),
\end{equation}
where $i$ is the color index. To attain the phenomenological side,
correlation function can be written as a complete set of
intermediate hadronic states with the same quantum numbers as the
current operator $J_{\mu}(x)$ can be inserted into the correlation
function. Then subtracting the ground state contribution from the
other quarkonium states and carrying out the integration over $x$,
we get:
\begin{eqnarray}\label{eq:Phys1}
\Pi _{\mu \nu }^{\mathrm{Phen.}}(q)&=&\frac{\langle 0|J_{\mu
}^{\chi_{c1}(1P)}|\chi_{c1}(1P)\rangle \langle
\chi_{c1}(1P)|J_{\nu }^{\dagger
\chi_{c1}(1P)}|0\rangle}{m_{\chi_{c1}(1P)}^{2}-q^{2}}  \nonumber \\
&+& \frac{\langle 0|J_{\mu }^{\chi_{c1}(2P)}|\chi_{c1}(2P)\rangle
\langle \chi_{c1}(2P)|J_{\nu }^{\dagger\chi_{c1}(2P)}|0\rangle
}{m_{\chi_{c1}(2P)}^{2}-q^{2}}+\ldots,
\end{eqnarray}
where $m_{\chi_{c1}(1P)}$ and $m_{\chi_{c1}(2P)}$ are the masses
of $\chi_{c1}(1P)$ and $\chi_{c1}(2P)$ states, respectively. The
dots in Eq.~(\ref{eq:Phys1}) imply contributions coming from
higher resonances and continuum states.

To complete the calculation of the phenomenological side of sum
rule we introduce the matrix elements through masses and decay
constants of $\chi_{c1}(1P)$ and its radial excited state
$\chi_{c1}(2P)$. The decay constants of $\chi_{c1}(1P,2P)$ which
is proportional to the matrix element of the axial current between
the one-P-meson state and the vacuum as:
\begin{eqnarray}\label{eq:CurDef.}
&&\langle0|J_{\mu}^{\chi_{c1}(1P)}|\chi_{c1}(1P)\rangle=f_{\chi_{c1}(1P)}m_{\chi_{c1}(1P)}\varepsilon_{\mu},
\end{eqnarray}
\begin{eqnarray}\label{eq:CurDef.}
&&\langle0|J_{\mu}^{\chi_{c1}(2P)}|\chi_{c1}(2P)\rangle=f_{\chi_{c1}(2P)}m_{\chi_{c1}(2P)}\widetilde{\varepsilon}_{\mu},
\end{eqnarray}
which can be considered as the overlap of quark and antiquark's
wave function. In Eq.~(\ref{eq:CurDef.}) $\varepsilon _{\mu }$ and
$\tilde{\varepsilon}_{\mu }$ are the polarization vectors of the
$\chi_{c1}(1P)$ and $\chi_{c1}(2P)$ mesons, respectively. Thus the
correlator $\Pi _{\mu \nu }^{\mathrm{Phen.}}(p)$ is defined by
\begin{eqnarray}\label{eq:Phys2}
\Pi_{\mu\nu}^{\mathrm{Phen.}}(q)&=&\frac{m_{\chi_{c1}(1P)}^{2}f_{\chi_{c1}(1P)}^{2}}{
m_{\chi_{c1}(1P)}^{2}-q^{2}}\Bigg( -g_{\mu \nu }+\frac{q_{\mu
}q_{\nu}}{m_{\chi_{c1}(1P)}^{2}}\Bigg)  \nonumber \\
&+&\frac{m_{\chi_{c1}(2P)}^{2}f_{\chi_{c1}(2P)}^{2}}{m_{\chi_{c1}(2P)}^{2}-q^{2}}\Bigg(-g_{\mu \nu }+\frac{%
q_{\mu }q_{\nu }}{m_{\chi_{c1}(2P)}^{2}}\Bigg) +\ldots.
\end{eqnarray}
Then the Borel transformation applied to Eq.~(\ref{eq:Phys2}) and
it yields
\begin{eqnarray}\label{eq:Phys3}
\mathcal{B}_{(q^2)}\Pi_{\mu\nu}^{\mathrm{Phen.}}(q^2)&=&m_{\chi_{c1}(1P)}^{2}f_{\chi_{c1}(1P)}^{2}e^{-m_{\chi_{c1}(1P)}^{2}/{M}^{2}}\times\left(-g_{\mu\nu}+\frac{q_{\mu}q_{\nu}}{m_{\chi_{c1}(1P)}^{2}}\right)  \nonumber \\
&+&m_{\chi_{c1}(2P)}^{2}f_{\chi_{c1}(2P)}^{2}e^{-m_{\chi_{c1}(2P)}^{2}/{M}^{2}}\times\left(-g_{\mu\nu}+\frac{q_{\mu}q_{\nu}}{m_{\chi_{c1}(2P)}^{2}}\right)\nonumber \\
&+&\ldots.
\end{eqnarray}
Here $M^{2}$ is the Borel mass parameter for the considered
states.

In the OPE side, the correlator can be stated as by contracting
the heavy quark fields in Eq.~(\ref{eq:CorrFunc}). After some
manipulations it reads
\begin{eqnarray}\label{eq:CorrFunc2}
\Pi _{\mu \nu }^{\mathrm{OPE}}(q)=-i\int d^{4}x~e^{iq\cdot
x}~\mathrm{Tr}~[{S}_{Q}^{ji}(-x)\gamma _{\mu }\gamma
_{5}S_{Q}^{ij}(x)\gamma _{5}\gamma _{\nu}],
\end{eqnarray}
where $S_c^{ij}$ is the heavy quark propagator and the explicit
form of it is given below ~\cite{Reinders}:
\begin{eqnarray}\label{eq:QProp}
S_{c}^{ij}(x)&=&i\int \frac{d^{4}k}{(2\pi )^{4}}e^{-ik\cdot x}\Bigg[ \frac{%
\delta _{ij}\Big( {\!\not\!{k}}+m_{c}\Big)
}{k^{2}-m_{c}^{2}}-\frac{gG_{ij}^{\alpha \beta }}{4}\frac{\sigma _{\alpha \beta }\Big( {%
\!\not\!{k}}+m_{c}\Big) +\Big(
{\!\not\!{k}}+m_{c}\Big)\sigma_{\alpha
\beta }}{(k^{2}-m_{c}^{2})^{2}}\nonumber \\
&+&\frac{g^{2}}{12}G_{\alpha \beta }^{A}G_{A}^{\alpha \beta
}\delta_{ij}m_{c}\frac{k^{2}+m_{c}{\!\not\!{k}}}{(k^{2}-m_{c}^{2})^{4}}+\ldots\Bigg].
\end{eqnarray}
In Eq.~(\ref{eq:QProp}) we use the following notations
\begin{eqnarray}\label{eq:Glu}
&&G_{ij}^{\alpha \beta }=G_{A}^{\alpha
\beta}t_{ij}^{A},\,\,~~G^{2}=G_{\alpha \beta }^{A}G_{\alpha \beta}^{A},  \nonumber\\
&&G^{3}=\,\,f^{ABC}G_{\mu \nu }^{A}G_{\nu \delta }^{B}G_{\delta
\mu }^{C},
\end{eqnarray}
with $A,B,C=1,\,2\,\ldots 8$ gluon color indices. In
Eq.~(\ref{eq:Glu}) $t^{A}=\lambda ^{A}/2$ with Gell-Mann matrices
$\lambda^{A}$, and the gluon field $G_{\alpha \beta}^{A}\equiv
G_{\alpha \beta }^{A}(0)$ is fixed at $x=0$.

The function $\Pi _{\mu \nu }^{\mathrm{OPE}}(q)$ has two different
structures and can be expressed as a sum of two components as
follows:
\begin{equation}
\Pi _{\mu \nu }^{\mathrm{OPE}}(q)=\Pi ^{\mathrm{OPE}}(q^{2})(-g_{\mu \nu })+%
\widetilde{\Pi }^{\mathrm{OPE}}(q^{2})q_{\mu }q_{\nu }.
\end{equation}
The QCDSR for the parameters of $\chi_{c1}(2P)$ can
be derived after equating the same structures in both $\Pi _{\mu \nu }^{\mathrm{Phen}%
}(q)$ and $\Pi _{\mu \nu }^{\mathrm{OPE}}(q)$. To continue our
evaluations, we select structure $(-g_{\mu \nu })$ at the later
stage. For Euclidean momentum $Q^2=-q^2>0$, the quantity $\Pi
^{\mathrm{OPE}}(q^{2})$ satisfies the dispersion relation as:
\begin{equation}\label{eq:PiQCD}
\Pi^{\mathrm{OPE}}(q^{2})=\int_{4m_{c}^{2}}^{\infty}ds~\frac{\rho
^{\mathrm{OPE}}(s)}{s-q^{2}}+\mathrm{subtracted~terms},
\end{equation}
where two-point spectral density is
$\rho^{OPE}_i(s)=\frac{1}{\pi}Im [\Pi_i(s)^{OPE}]$ as $i$
represent operator dimensions.
\begin{equation}\label{eq:rho}
\rho^{\mathrm{OPE}}(s)=\rho^\mathrm{Pert}(s)+\rho^\mathrm{(4)}(s)+\rho^\mathrm{(6)}(s)+\rho^\mathrm{(8)}(s).
\end{equation}
Concrete expressions of the spectral densities are given in
Appendix. After assuming the quark-hadron duality and applying
Borel transform to subtract the contribution of the higher
resonances and continuum states, at the end the sum rules for
$\chi_{c1}(1P)$ state is found as follows:
\begin{equation}\label{eq:MassSR1P}
m_{\chi_{c1}(1P)}^{2}=\frac{\int_{4m_{c}^{2}}^{s_{0}}ds~\rho ^{\mathrm{OPE}%
}(s)~s~e^{-s/M^{2}}}{\int_{4m_{c}^{2}}^{s_{0}}ds~\rho^{%
\mathrm{OPE}}(s)~e^{-s/M^{2}}},
\end{equation}
\begin{equation}\label{eq:DecConSR1P}
f_{\chi_{c1}(1P)}^{2}=\frac{1}{m_{\chi_{c1}(1P)}^{2}}\int_{4m_{c}^{2}}^{s_{0}}ds~\rho
^{\mathrm{OPE}}(s)e^{\Big(m_{\chi_{c1}(1P)}^{2}-s\Big)/M^{2}}.
\end{equation}
In the above expressions $M^{2}$ is the Borel mass parameter and
$s_0$ is the continuum threshold, which separates the contribution
of the ground state $\chi_{c1}(1P)$ from the higher resonances and
continuum. As for the $\chi_{c1}(2P)$ resonance we achieve the sum
rules as:
\begin{equation}\label{eq:MassSR2P}
m^{2}_{\chi_{c1}(2P)}=\frac{\int_{4m_{c}^{2}}^{s_{0}^{\ast}}ds~\rho^{\mathrm{OPE}
}(s)~s~e^{-s/M^{2}}-f_{\chi_{c1}(1P)}^{2}m_{\chi_{c1}(1P)}^{4}e^{-m_{\chi_{c1}(1P)}^{2}/M^{2}}}{\int_{4m_{c}^{2}}^{s_{0}^{\ast
}}ds~\rho^{\mathrm{OPE}}(s)~e^{-s/M^{2}}-f_{\chi_{c1}(1P)}^{2}m_{\chi_{c1}(1P)}^{2}e^{-m_{\chi_{c1}(1P)}^{2}/M^{2}}},
\end{equation}
\begin{eqnarray}\label{eq:DecConSR2P}
f_{\chi_{c1}(2P)}^{2}&=&\frac{1}{m_{\chi_{c1}(2P)}^{2}}\Bigg[
\int_{4m_{c}^{2}}^{s_{0}^{\ast}}ds~\rho
^{\mathrm{OPE}}(s)~e^{\Big(m_{\chi_{c1}(2 P)}^{2}-s\Big)/M^{2}}\nonumber\\
&-&f_{\chi_{c1}(1P)}^{2}m_{\chi_{c1}(1P)}^{2}e^{\Big(m_{\chi_{c1}(2P)}^{2}-m_{\chi_{c1}(1P)}^{2}\Big)/M^{2}}\Bigg]
\end{eqnarray}
where $s_{0}^{\ast }$ is the continuum threshold parameter, which
separates the contribution of the
``$\chi_{c1}(1P)+\chi_{c1}(2P)$'' from the ``higher resonances and
continuum''. As we know that sum rules rely on the same spectral
density $\rho ^{\mathrm{QCD}}(s)$ and the continuum threshold has
to obey $s_0 < s_0^{\star}$. We have pointed out above the mass
and decay constant of $\chi_{c1}(1P)$ entering into
Eqs.~(\ref{eq:MassSR2P}) and ~(\ref{eq:DecConSR2P}) as the input
parameters (For similar works see
Refs.~\cite{Azizi:2017izn,Agaev:2017tzv,Agaev:2017lip}).

\section{Numerical Analysis}\label{sec:NumAnal}

To perform and continue the numerical analysis for the studied
states, we employed the input parameter values in
Table~\ref{tab:inputPar} in computations:
\begin{table}[htbp]
\caption{\label{tab:inputPar}Input
parameters~\cite{Tanabashi,Narison:2010cg}.}
\begin{indented}
\item[]\begin{tabular}{@{}ll} \br
Parameters &Values\\
\mr
$m_{c}$                                     & $(1.67\pm0.07)~\mathrm{GeV}$ \\
$m_{\chi_{c1}(1P)}$                         & $(3510.67 \pm0.05)~\mathrm{MeV}$ \\
$\langle\frac{\alpha_sG^2}{\pi}\rangle $    & $(0.012\pm0.004)~\mathrm{GeV}^4 $\\
$\langle g_{s}^3G^3\rangle$                 & $(0.57\pm0.29)~\mathrm{GeV}^6 $\\
\br
\end{tabular}
\end{indented}
\end{table}

The QCDSR obtained in this work allows us to calculate
characteristics of the axial-vector ground-state and its first
radial excited state of $\chi_{c1}$. These characteristic
quantities depend on the Borel mass parameter $M^2$ and continuum
threshold $s_0$. The continuum threshold $s_0$ is not completely
arbitrary, since it is correlated to the energy of the first
exited state. Nevertheless, the dependence of mass and decay
constant sum rules on these parameters should remain within the
acceptable limits.

The choice of arbitrary parameters $M^{2}$ and $s_{0}$ has to
satisfy standard restrictions. The parameters $s_0$ and
$s_{0}^{*}$ are defined from the conditions that guarantee the sum
rules to have the best stability in the allowed $M^{2}$ regions.
For the greatest accessible values of $M^{2}$ the perturbative
contribution has to constitute more than $50\%$ part of the total
contribution. As concerns the lower bound of $M^{2}$, the
non-perturbative contribution of any dimension should include at
most $\sim20\%$ part of the full contribution. Boundaries of
$M^{2}$ are fixed by analyzing the pole contribution. Minimal
dependence of the extracted quantities on $M^{2}$ while varying
$s_{0}$ is another constraint that has to be imposed. Consequently
performed analysis leads to the following working windows to the
$M^{2}$ and $s_{0}$ for the state ${\chi_{c1}(1P)}$:
\begin{eqnarray*}
\quad\quad\quad\quad\quad\quad 3~\mathrm{GeV}^2 \leq M^2 \leq 6~\mathrm{GeV}^2 \nonumber\\
\quad\quad\quad\quad\quad\quad 13.03~\mathrm{GeV}^2 \leq s_0 \leq 14.52~\mathrm{GeV}^2. \nonumber\\
\end{eqnarray*}
Also, we choose the regions for the Borel mass parameter and
continuum threshold for the resonance ${\chi_{c1}(2P)}$. Our
numerical result is point out the following interval to us:
\begin{eqnarray*}
\quad\quad\quad\quad\quad\quad 3~\mathrm{GeV}^2 \leq M^2 \leq 6~\mathrm{GeV}^2 \nonumber\\
\quad\quad\quad\quad\quad\quad 16.40~\mathrm{GeV}^2 \leq s_0 \leq 18.06~\mathrm{GeV}^2 \nonumber\\
\end{eqnarray*}
Then, numerical results of the calculations are gathered in Table
\ref{tab:MassResults} and \ref{tab:DecConResults}, where we
present the mass and decay constants of the $\chi_{c1}(2P)$ and
$\chi_{c1}(1P)$ mesons (Only the references in recent years are
given in the tables). For the ground-state $\chi_{c1}(1P)$ we
found the mass value as $3554\pm        \mathrm{MeV}$. It is seen
that $m_{\chi_{c1}(1P)}$ is roughly in agreement with the
experimental data, but within the error limit of the calculations
of the QCDSR.

\begin{figure}[htbp]
\begin{center}
\includegraphics[totalheight=5cm,width=7cm]{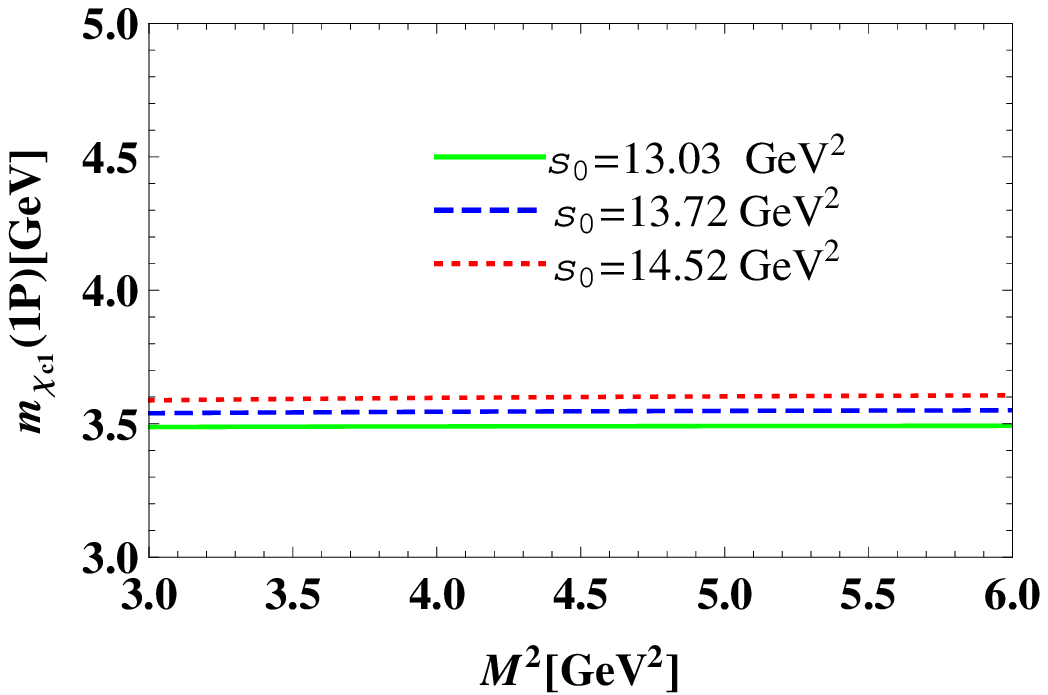}\hskip1cm
\includegraphics[totalheight=5cm,width=7cm]{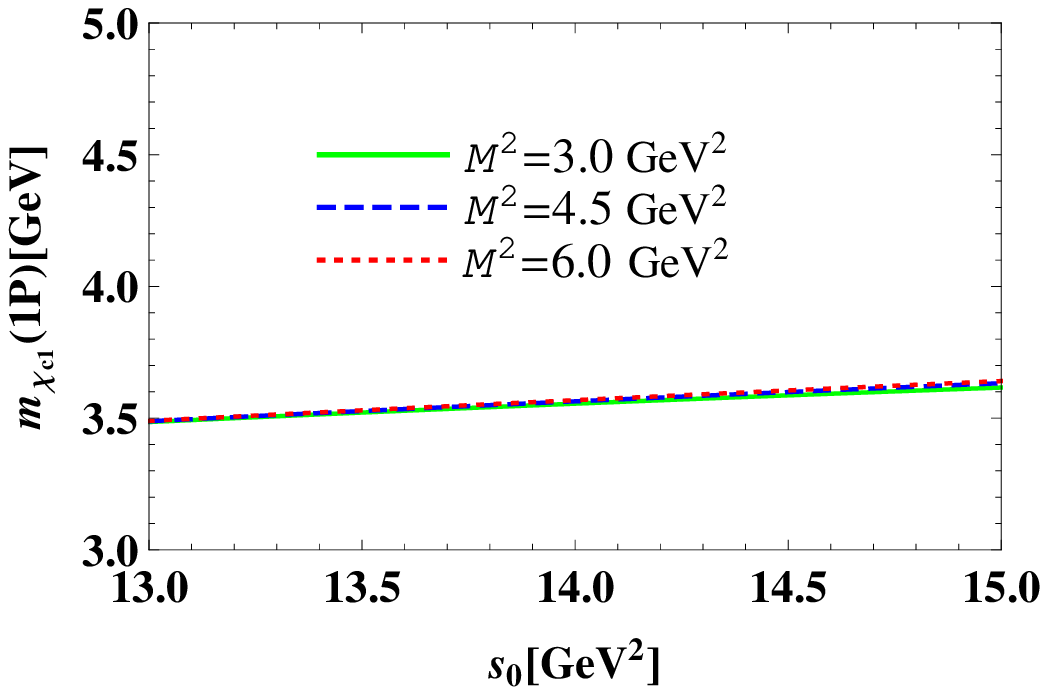}\hskip1cm
\includegraphics[totalheight=5cm,width=7cm]{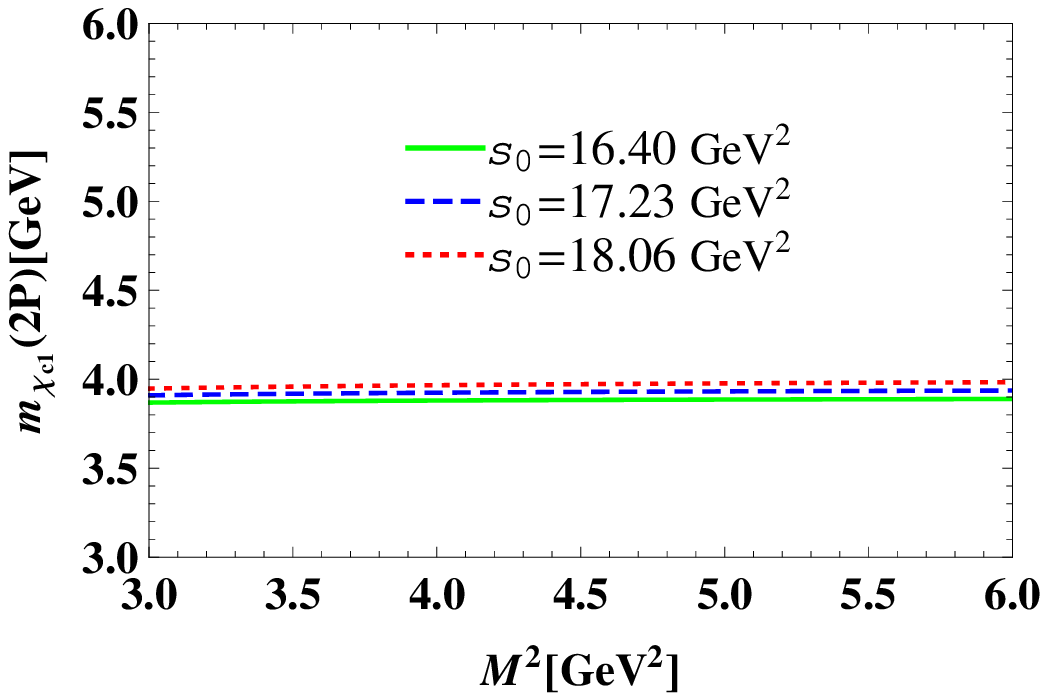}\hskip1cm
\includegraphics[totalheight=5cm,width=7cm]{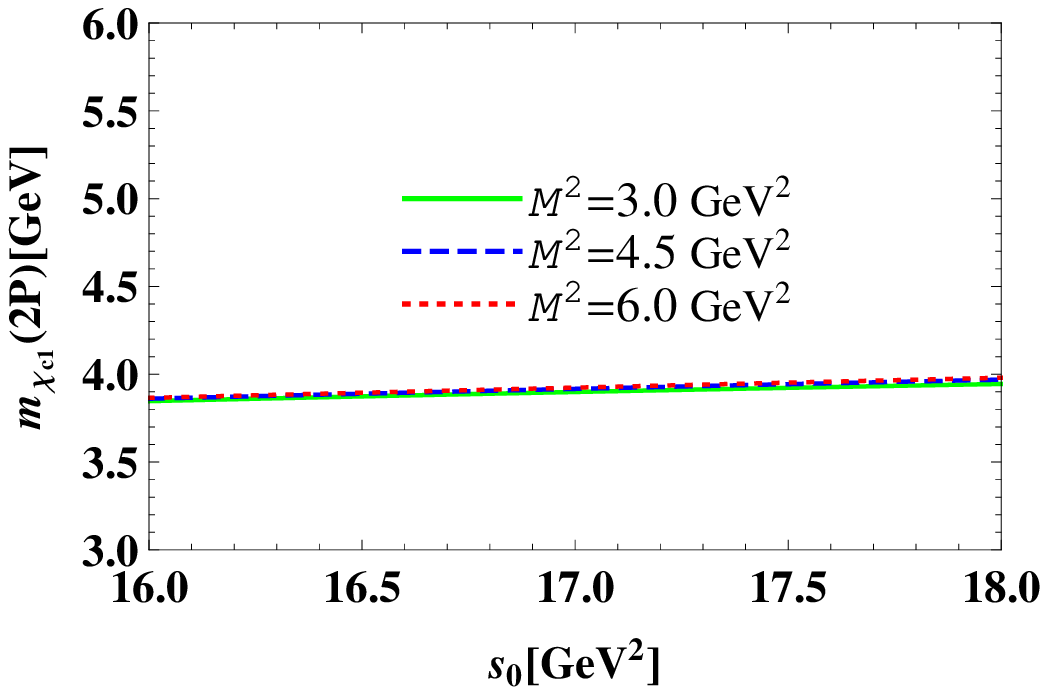}
\caption{ (Left panel) The mass of the meson
$m_{\chi_{c1}(1P,2P)}$ as a function of the Borel parameter $M^2$
for fixed $s_0$, and (Right panel) the continuum threshold $s_0$
for fixed $M^2$.} \label{fig:mass}
\end{center}
\end{figure}

\begin{figure}[htbp]
\begin{center}
\includegraphics[totalheight=5cm,width=7cm]{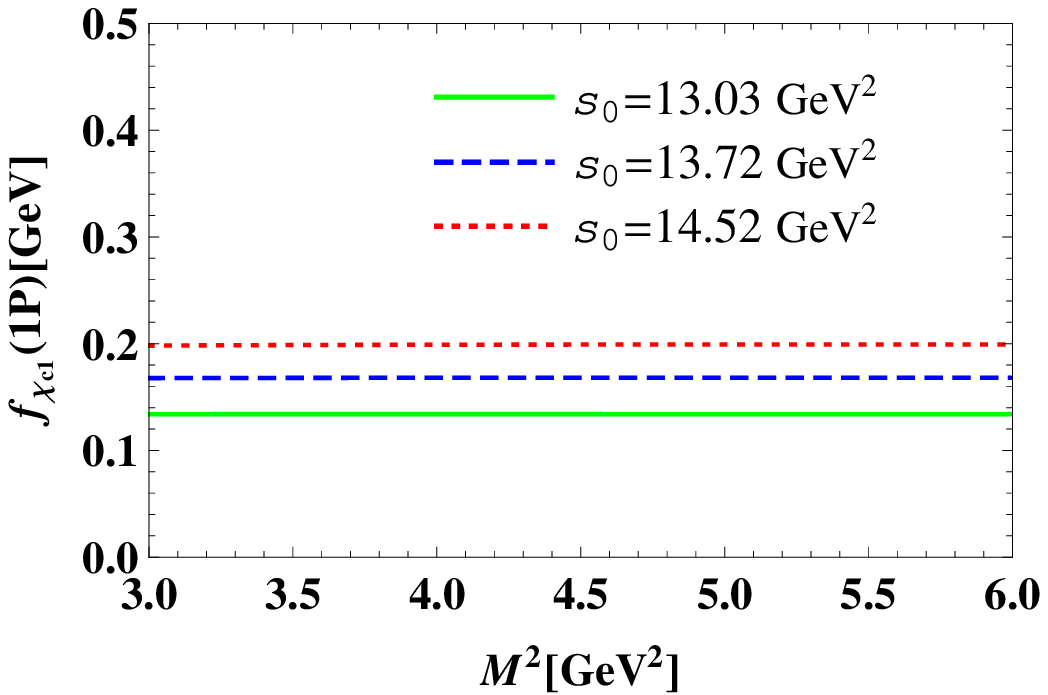}\hskip1cm
\includegraphics[totalheight=5cm,width=7cm]{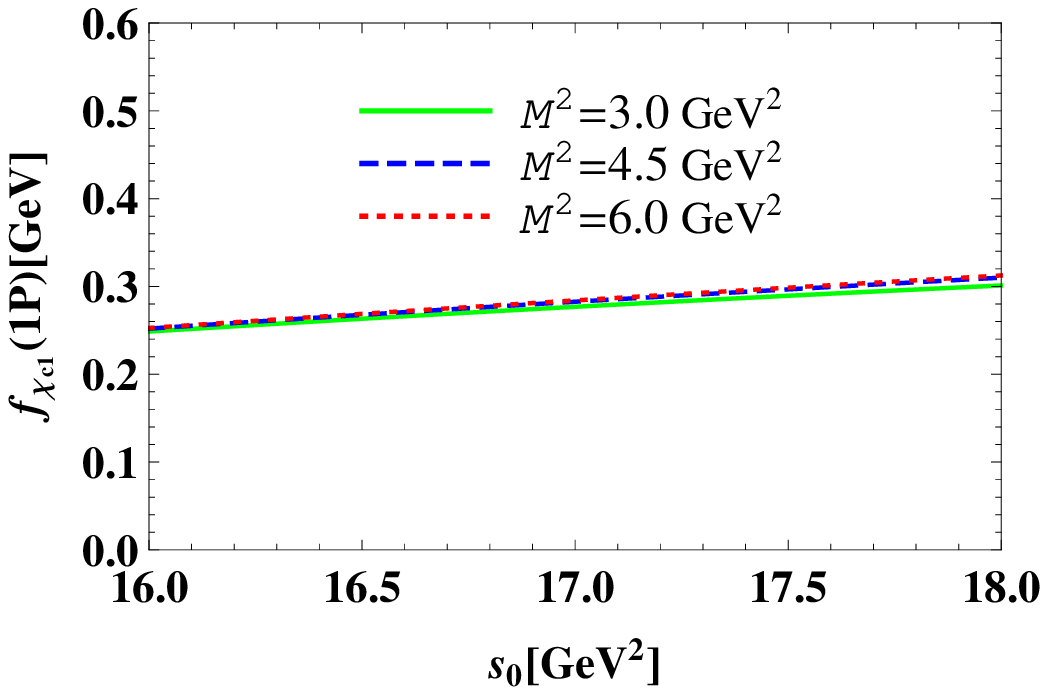}\hskip1cm
\includegraphics[totalheight=5cm,width=7cm]{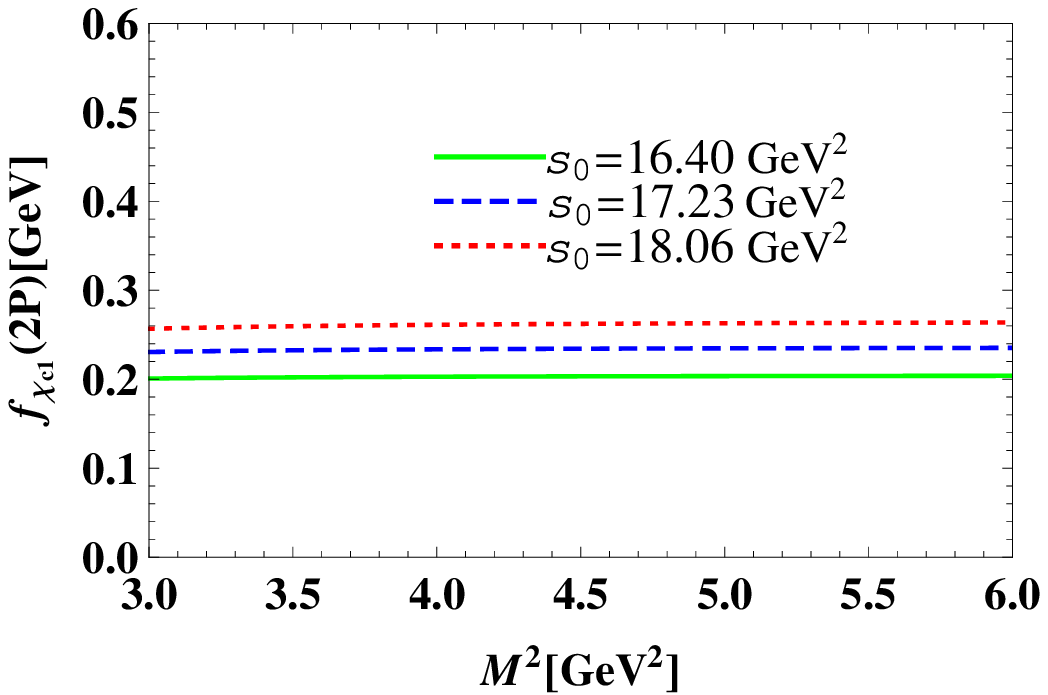}\hskip1cm
\includegraphics[totalheight=5cm,width=7cm]{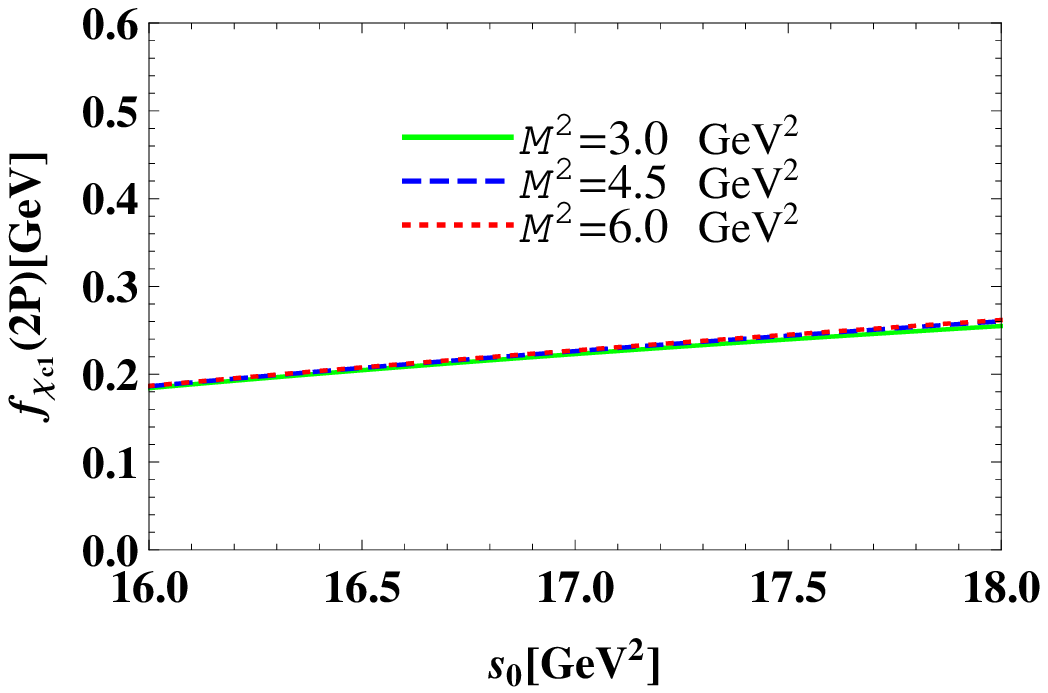}
\caption{(Left panel) The dependence of the decay constant
$f_{\chi_{c1}(1P,2P)}$ on the Borel mass parameter at chosen
values of $s_0$, and (Right panel) on the $s_0$ for fixed $M^2$.}
\label{fig:DecCon}
\end{center}
\end{figure}

In the following drawn figure \ref{fig:mass} and \ref{fig:DecCon},
we show the dependence of $m_{\chi_{c1}(1P)}$,
$f_{\chi_{c1}(1P)}$, $m_{\chi_{c1}(2P)}$ and $f_{\chi_{c1}(2P)}$
on $M^2$ at fixed $s_0$ and as function of $s_0$ for chosen values
of $M^2$. As one can see, the mass of the $\chi_{c1}(1P)$ meson is
rather stable against variations at $M^2$
and $s_{0}^{*}$. Additionally, theoretical errors for $f_{\chi_{c1}(1P)}$ and $f_{\chi_{c1}(2P)}$ arising from uncertainties of $%
M^2$ and $s_0$ and other input parameters remain within allowed
territory for theoretical errors hereditary in the sum rule
computations which is acceptable up to roughly $20\%$ of
predictions.
\begin{table}[htbp]
\caption{The $\chi_{c1}(2P)$ meson masses from different models.}
\label{tab:MassResults}
\lineup
\begin{center}
\begin{tabular}{cccc}
\hline
Mass $[\mathrm{MeV}]$               &$m_{\chi_{c1}(1P)}$                                          & $m_{\chi_{c1}(2P)}$               \\
\hline
$\mathrm{Experiment}$               &~~~~~~~~~~~~~~~~$3510.67 \pm 0.05$~\cite{Tanabashi}          & -                                  \\
$\mathrm{Our~Work}$                 &~~~$3545^{+62}_{-57}$                                        &~~$3924^{+48}_{-56}$                  \\
QCD Sum Rules                       &~~~~~$3520$~\cite{VeliVeliev:2012cc}                         & -                                  \\
                                    &~~~~~$3510$~\cite{Palameta:2018yce}                          & -                                   \\
Covariant Bethe-Salpeter            &~~~~~$3437$~\cite{Blank:2011ha}                              & -                                   \\
Equation Approach                   &                                                             &                                      \\
Quark Model                         &    -                                                        & ~~~~$3950$~\cite{Godfrey:1985xj}     \\
Regge Trajectories                  &~~~~~~$3511$~\cite{Ebert:2011jc}                             & ~~~~~~$3906$~\cite{Ebert:2011jc}     \\
Modified Regge trajectory           &    -                                                        & ~~~~~~$3922$~\cite{Sonnenschein:2016ibx}\\
Potential Model                     &    -                                                        & ~~~~~~$3950$~\cite{Radford:2007vd}    \\
QCD-inspired Quark Potential Model  &    -                                                        & ~~~~~~$3934$~\cite{Cao:2012du}    \\
using Gaussian Expansion Method     &                                                             &                                        \\
Constituent Quark Model             &    -                                                        & ~~~~~~$3947$~\cite{Segovia:2013wma}   \\
Friedrichs-Model-like Scheme        &    -                                                        & ~~~~~$3920$~\cite{Zhou:2017dwj}        \\
Non-relativistic Quark Model        &~~~~~$3510$~\cite{DSouza:2017hno}                            & ~~~~~~$3872$~\cite{DSouza:2017hno}       \\
\hline
\end{tabular}
\end{center}
\end{table}
Our results for the decay constants of corresponding mesons are
presented in Table \ref{tab:DecConResults}.
\begin{table}[htbp]
\caption{The decay constants of the $\chi_{c1}(1P)$ and
$\chi_{c1}(2P)$ mesons.} \label{tab:DecConResults}
\begin{center}
\begin{tabular}{ccccc}
\hline
Decay constant                                      & $f_{\chi_{c1}(1P)}[\mathrm{MeV}]$                  & $f_{\chi_{c1}(2P)}[\mathrm{MeV}]$  \\
\hline
Our~Work                                            & ~~~~~$167^{+33}_{-33}$                             & $225^{+32}_{-33}$                   \\
QCD Sum Rules                                       & ~~~~~~~~~~~~~~$344\pm27$~\cite{VeliVeliev:2012cc}  &  -                       \\
                                                    & ~~~~~~~$185$~\cite{Rui:2017pre}                    &  -                       \\
Non-relativistic QCD                                & ~~~~~~~$140$~\cite{Luchinsky:2018dwk}              &  -                       \\
Factorization Approach                              & ~~~~~~~~~~~~~~$295\pm28$~\cite{Munoz:2010xv}       &  -                       \\
\hline
\end{tabular}
\end{center}
\end{table}

\section{Concluding Remarks}\label{sec:Result}

We can summarize the present work by stating that a study of the
$\chi_{c1}(1P)$ and $\chi_{c1}(2P)$ states has been carried out by
employing QCD sum rules method, where in calculations terms up to
dimension-8 have been computed. We adopted interpolating currents
for $\chi_{c1}(1P)$ and $\chi_{c1}(2P)$ charmonia with quantum
numbers $J^{PC}=1^{++}$. The mass and decay constant of the
ground-state meson $\chi_{c1}(1P)$ and its first radial excitation
$\chi_{c1}(2P)$ have been extracted from the corresponding QCDSR.

As is seen, our results obtained for the mass of $\chi_{c1}(2P)$
state by treating it as a first excited state of $\chi_{c1}(1P)$
is smaller than the mass of the $\chi_{c1}(4140)$ with a
$~16.78\%$ difference. If we compare it to the $X(3872)$ particle,
there is a mass difference of about $~9.22\%$. However  the
dominant idea on these particles in the literature that $X(3872)$
and $\chi_{c1}(4140)$ are probably exotic particles.  Our mass
value which we found as a result of calculations is very close to
the $X(3940)$ particle. But we still don't know the quantum
numbers of this state. Also the QCD sum rule predictions for the
mass and decay constants extracted in the present work by handling
the interpolating current suffer from the large uncertainties.
Anyhow such errors are inherent in the sum rule calculations, and
are inevitable part of the whole picture. Therefore precise
determination of the fundamental properties of charmonia is very
important to explain the differences between experimental data and
theoretical predictions. We hope that the theoretical studies and
more sensitive experimental data will clarify our knowledge on
this issue. To determine these hadronic parameters is also an
important issue for the completion of the hadron spectrum. So
these study can provide information and clues on the
identification of the XYZ mesons. Our results may favorable in
resolving the long-standing puzzle of determining the observed
P-wave state, and also the interpretation of the enigmatic
$X(3872)$ and $\chi_{c1}(4140)$ state \cite{Zhou:2017dwj}.

Finally the charmonia are essential research area both at the
running and projected in many large-scale experiments such as
Belle, BESIII, LHC, BaBar and FAIR. The upcoming high precision
data from BESIII, LHCb, Belle and BaBar as well as from the future
detectors BelleII and PANDA will allow us to deeply understand the
spectrum of the excited states and also the nature of the exotics.

\appendix

\section{Spectral Density}\label{sec:App}

The last explicit form of the perturbative part of the spectral
density Eq.~(\ref{eq:rho})
\begin{equation}\label{eq:rhopert}
\rho^{\mathrm{Pert.}}(s)=\frac{1}{\pi^2}\Bigg(\frac{m_c^2}{s}+\frac{1}{8}\Bigg)\sqrt{s(s-4m_c^2)}
\end{equation}
The nonperturbative part of the spectral density Eq.~
(\ref{eq:rho}) is determined by the formula corresponding to the
dimension four ($\rho_{4}$), six ($\rho_{6}$) and eight
($\rho_{8}$), respectively:
\begin{eqnarray}\label{eq:NPert}
\rho ^{\mathrm{Nonpert.}}(s)&=&\Big \langle\frac{\alpha
_{s}G^{2}}{\pi }\Big\rangle \int_{0}^{1}~\rho^\mathrm{(4)}(z,s)~dz  \nonumber \\
&+&\Big \langle g_{s}^{3}G^{3}\Big \rangle\int_{0}^{1}~\rho^\mathrm{(6)}(z,s)~dz  \nonumber \\
&+&\Big \langle\frac{\alpha _{s}G^{2}}{\pi }\Big \rangle^{2}%
\int_{0}^{1}~\rho^\mathrm{(8)}(z,s)~dz.
\end{eqnarray}
In Eq.~(\ref{eq:NPert}) the functions $~\rho_{4}(z,s),\
~\rho_{6}(z,s)$ and $~\rho_{8}(z,s)$ have the explicit forms:
\begin{eqnarray}
~\rho^{\mathrm{(4)}}(z,s)&=&\frac{1}{24r^2}\Big[3r^2(s-\varphi)+[m_c^2(3+9r)+sr^2] \nonumber \\
&\times&\delta^{(1)}(s-\varphi)+ m_c^2 s (1 -2z)^2
\delta^{(2)}(s-\varphi)\Big], \nonumber \\
\end{eqnarray}
\begin{eqnarray}
~\rho^{\mathrm{(6)}}(z,s)&=&\frac{1}{15\cdot 2^{9}\pi^2 r^5}\Bigg\{12\delta^{(1)}(s-\varphi)r^3 [1+5 r(1 + r)] \nonumber \\
&+& 2r^2 \delta^{(2)}(s-\varphi)\Big[27 m_c^2[1 + 5 r (1 + r)]+ 2s r \nonumber \\
&\times&[3 +r (19+ 27 r)]\Big]+r \delta^{(3)}(s-\varphi)\Big[-15 m_c^4 \nonumber \\
&\times&[1 + 5 r (1 + r)]+ 6 m_c^2 s r [3+ r (23+39r)]\nonumber \\
&+&s^2 r^3 (8 + 27 r)]\Big]-\delta^{(4)} (s-\varphi)\Big[2 m_c^6[1 + 5 r \nonumber \\
&\times& (1 + r)]+ m_c^4 s r [7 + r (39+47r)]-r^5 s^3\nonumber \\
&-&6 m_c^2 r^3 s^2(1-2z)^2\Big]\Bigg\}
\end{eqnarray}
and
\begin{eqnarray}
~\rho^{\mathrm{(8)}}(z,s)&=&-\frac{1}{2^4\cdot3^3 r^{2}}
m_c^2 \pi^2\Big[6r \delta^{(3)}(s-\varphi)+ \delta^{(5)}(s-\varphi) \nonumber \\
&\times& s (m_c^2 + r s)+\delta^{(4)}(s-\varphi)[m_c^2 + 2 s (1 +3r)]\Big] \nonumber \\
\end{eqnarray}
where we use the following notations
\begin{equation}
\ r=z(z-1),\ \ \varphi =\frac{m_{c}^{2}}{z(1-z)}.
\end{equation}
In the above expressions  the Dirac delta function $\ \delta
^{(n)}(s-\varphi )$ is defined as
\begin{equation}
\delta^{(n)}(s-\varphi)=\frac{d^{n}}{ds^{n}}\delta (s-\varphi).
\end{equation}

\section*{Acknowledgment}\label{sec:e}

Authors thank to E. Veli Veliev for enlightening and helpful
discussions and also Kocaeli University for the partial financial
support through the grant BAP 2018/082.

\end{document}